# Improving Josephson junction reproducibility for superconducting quantum circuits: shadow evaporation and oxidation

Dmitry O. Moskalev[1,2], Evgeniy V. Zikiy[1,2], Anastasiya A. Pishchimova[1,2], Daria A. Ezenkova[1,2], Nikita S. Smirnov[1], Anton I. Ivanov[1], Nikita D. Korshakov[1], and Ilya A. Rodionov[1,2,*]

[1] FMN Laboratory, Bauman Moscow State Technical University, Moscow, 105005, Russia
[2] Dukhov Automatics Research Institute (VNIIA), Moscow, 127055, Russia
*Electronic mail: irodionov@bmstu.ru

**ABSTRACT**

The most commonly used physical realization of superconducting qubits for quantum circuits is a transmon. There are a number of superconducting quantum circuits applications, where Josephson junction critical current reproducibility over a chip is crucial. Here, we report on a robust chip scale $Al/AlO_x/Al$ junctions fabrication method due to comprehensive study of shadow evaporation and oxidation steps. We experimentally demonstrate the evidence of optimal Josephson junction electrodes thickness, deposition rate and deposition angle, which ensure minimal electrode surface and line edge roughness. The influence of oxidation method, pressure and time on critical current reproducibility is determined. With the proposed method we demonstrate $Al/AlO_x/Al$ junction fabrication with the critical current variation ($\sigma/\langle I_c \rangle$) less than 3.9 % (from 150×200 to 150×600 $nm^2$ area) and 7.7 % (for 100×100 $nm^2$ area) over 20×20 $mm^2$ chip. Finally, we fabricate separately three 5×10 $mm^2$ chips with 18 transmon qubits (near 4.3 GHz frequency) showing less than 1.9 % frequency variation between qubits on different chips. The proposed approach and optimization criteria can be utilized for a robust wafer-scale superconducting qubit circuits fabrication.

**Introduction**

Superconducting quantum circuits platform is a promising solution for logic quantum gates realization and quantum simulators [1-6]. The key element of superconducting quantum circuits is a three-layer nanoscale aluminum Josephson junction (JJ). Over the past two decades, a number of methods is proposed to either improve Josephson junction critical current reproducibility [7, 8] or replace shadow evaporation [9], but it is still on a roll. Josephson junction area variation is the dominant reason of junction critical current variation resulting in qubit frequencies instability, which is crucial for multiple qubit systems. Additionally, high critical current reproducibility ensures proper qubit frequencies calculation for subsequent control without cross-talks [10]. Moreover, JJs critical current fluctuation is of a great importance for traveling-wave parametric amplifiers (TWPA), since it can lead to multiple reflections from thousands of junctions and loss along lumped-element transmission line [11, 12].

Transmon is one of the most widespread superconducting qubit types as it less sensitive to charge noise, has better scalability and control compared to other qubits [7]. Recently, for transmon-like qubit fabrication technology JJs critical current variations of 3.5 % for JJ area of 0.042 $um^2$ over 49 $cm^2$ chip is demonstrated [7]. Another paper also shows 3.7 % critical current variation over a wafer that contains forty 0.5×0.5 $cm^2$ chips with Josephson junction area ranging from 0.01 to 0.16 $um^2$ [8]. However, there are no experimental results in the papers linking JJ reproducibility to Al thin-film electrodes and $AlO_x$ oxide properties.

In this paper, we demonstrate a robust chip scale $Al/AlO_x/Al$ junction fabrication method based on a comprehensive study of shadow evaporation and oxidation steps. We examined a lower electrode deposition stage through two-layer resist mask and revealed the process parameters that determine its thickness variation, surface morphology and line edge roughness. Next, we showed the effect of the shadow evaporation stage on the linear critical dimensions of electrodes, which directly affects the JJ critical current reproducibility. Then we investigated tunnel barrier formation stage (oxidation) and optimized oxidation process parameters to improve the reproducibility of sub-200 nm $Al/AlO_x/Al$ Josephson junctions. In order to characterize JJs reproducibility we fabricated statistically significant number (more than 2500) of Josephson junctions and directly measuring its normal resistance variation. We demonstrate the critical current variation ($\sigma/\langle I_c \rangle$) less than 3.9 % (from 150×200 to 150×600 $nm^2$ area) and 7.7 % (for 100×100 $nm^2$ area) over 20×20 $mm^2$ chip. Finally, we fabricate separately three 5×10 $mm^2$ chips with 18 transmon qubits (near 4.3 GHz frequency) showing less than 1.9 % frequency variation between qubits on different chips.

## Experimental details

For this study, we used high-resistivity silicon substrates. Prior to the base layer deposition, the substrate is cleaned in a Piranha solution at 80°C, followed by dipping in hydrofluoric bath [13]. 100 nm thick Al base layer is deposited using ultra high vacuum e-beam evaporation system. Pads were defined using a direct-laser lithography and dry-etched in $BCl_3/Cl_2$ inductively coupled plasma. The Josephson junctions were fabricated using Dolan technique [14]. The substrate is spin coated with resist bilayer composed of 500 nm EL9 copolymer and 100 nm CSAR 62. Layouts were generated and exposed with 50 keV e-beam lithography system. The development was performed in a bath of amylacetate followed by IPA dip and additional in a IPA:DIW solution. $Al/AlO_x/Al$ junctions are shadow evaporated in ultra-high vacuum deposition system. Resist lift-off was performed in N-methyl-2-pyrrolidone at 80°C for 3 hours. Finally, we patterned and evaporated aluminum bandages using the same process as for junctions with an in-situ Ar ion milling.

The Josephson junctions room temperature resistance were individually measured with automated probe station. Some of JJs were also measured in a cryogen-free cryostat to confirm critical current evaluation. The quality and uniformity of the deposited electrodes was examined using a scanning electron microscopy.

## Experimental results and discussion

**Shadow evaporation of Josephson junctions.** During electron-beam evaporation of Josephson junctions a diverging metal flow is formed (Fig. 1a). Its conical shape leads to nanoscale misalignment of junction electrode positions over a wafer and changes in electrode linear dimensions. One can imagine this effect as changing deposition angle over the wafer - from the center point of the wafer to edges. The changing angle results in a different shading of a metal flow from a two-layer resist mask edge. Additionally, it also affects thickness fluctuations of shadow evaporated aluminum electrodes which can change its band gap [15], and, therefore, the critical current of the Josephson junction.

We demonstrated that JJ electrodes thickness nonuniformity becomes higher as deposition angle increases (up to 14% over the 4-inch wafer at 60° deposition angle) which well compatible with the simulation results (Fig. 1c,d). We observed reduction of 100-nm width electrode linear dimensions of 18 % over the 4-inch wafer (Fig. 1b). The electrode dimensions reduction is symmetrical against the central line of the wafer. These effects become much more significant for wafer-scale fabrication and for larger area superconducting devices with long Josephson junction arrays. To reduce the effects of shadow evaporation stage on linear dimensions and thickness variation, we placed all our chips in position B (Fig. 1b) in the center of the substrate holder for this study.

Previous study shown that only 10 % of AlOx tunnel barrier area of Josephson junction is actively participate in a tunneling process [16]. AlOx thickness variations are dominantly caused by grain boundary grooving in a bottom polycrystalline Al-electrode [17, 18]. There are three main stages (Fig. 2a) of electrodes shadow evaporation at an angle to a surface [19]. First, individual vapor species arrive at random locations on the surface with a given tilt angle. Second, deposited particles accumulate within certain regions in the form of grains of material which start casting shadows to neighbor surface zones, where vapor species cannot be deposited. At the third stage, taller surface features are more likely to grow, initiating a competitive growth process where the taller the feature the larger its shadow. Microstructure and surface of the bottom Al electrode mainly determine the properties of the Al/AlO$_x$/Al system and have to be optimized, as it forms the base for the subsequent tunnel barrier growth.

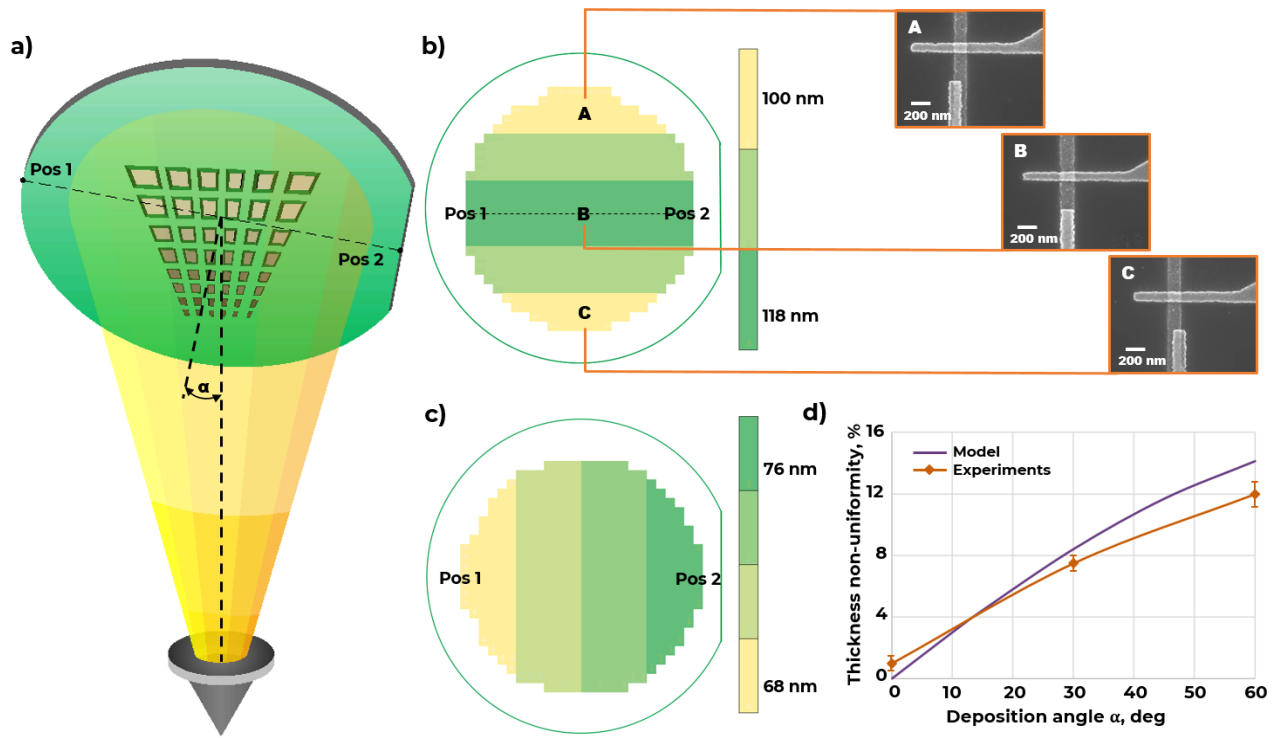

**Figure 1.** Thin film evaporation diagram from crucible at an angle to a substrate (a); distribution map of JJ dimensions over a 4" wafer (α = 0°) (b); distribution map of the film thickness over a 4" wafer (α = 60°) (c); the dependence of a film thickness nonuniformity vs deposition angle α (d).

There are three key parameters of e-beam evaporation process affecting the microstructure and surface quality of thin-film electrodes: deposition rate, electrode thickness and deposition angle. Here, we investigate e-beam evaporation of bottom Al electrodes with deposition rate in the range of 0.2–1.5 nm/s, electrode thickness from 15 to 45 nm (the lower limit was determined by the film continuity condition at room temperature), and deposition angle from 0° to 60°. Fig. 2b,c show an extremum that corresponds to minimum RMS surface roughness and line edge roughness (LER) of the bottom electrode. This dependence is typical for a wide range of deposition angles (from 0° to 60°). Deposition rate plays an important role in controlling adatoms surface diffusion and relocation. Rapid arrival of new particles (adatoms and cluster) could bury and interfere with particles diffusing on the substrate surface or thin film surface. Lower deposition rates allow ad-particles to diffuse over greater lengths.

Since RMS surface roughness of thin aluminum films increases with increasing thickness [20], in our work we investigated a combined effect from deposition angle and film thickness to bottom electrode surface and LER. We experimentally shown that decreased deposition angle contributes to the minimization of RMS surface roughness (Fig. 2d) and electrode line edge roughness (Fig. 2e). According to the model of film growth at angle, the shading effect becomes more significant as the deposition angle increases [21]. Additionally, increased film thickness leads to surface roughness increases due to the competitive growth of individual crystallites.

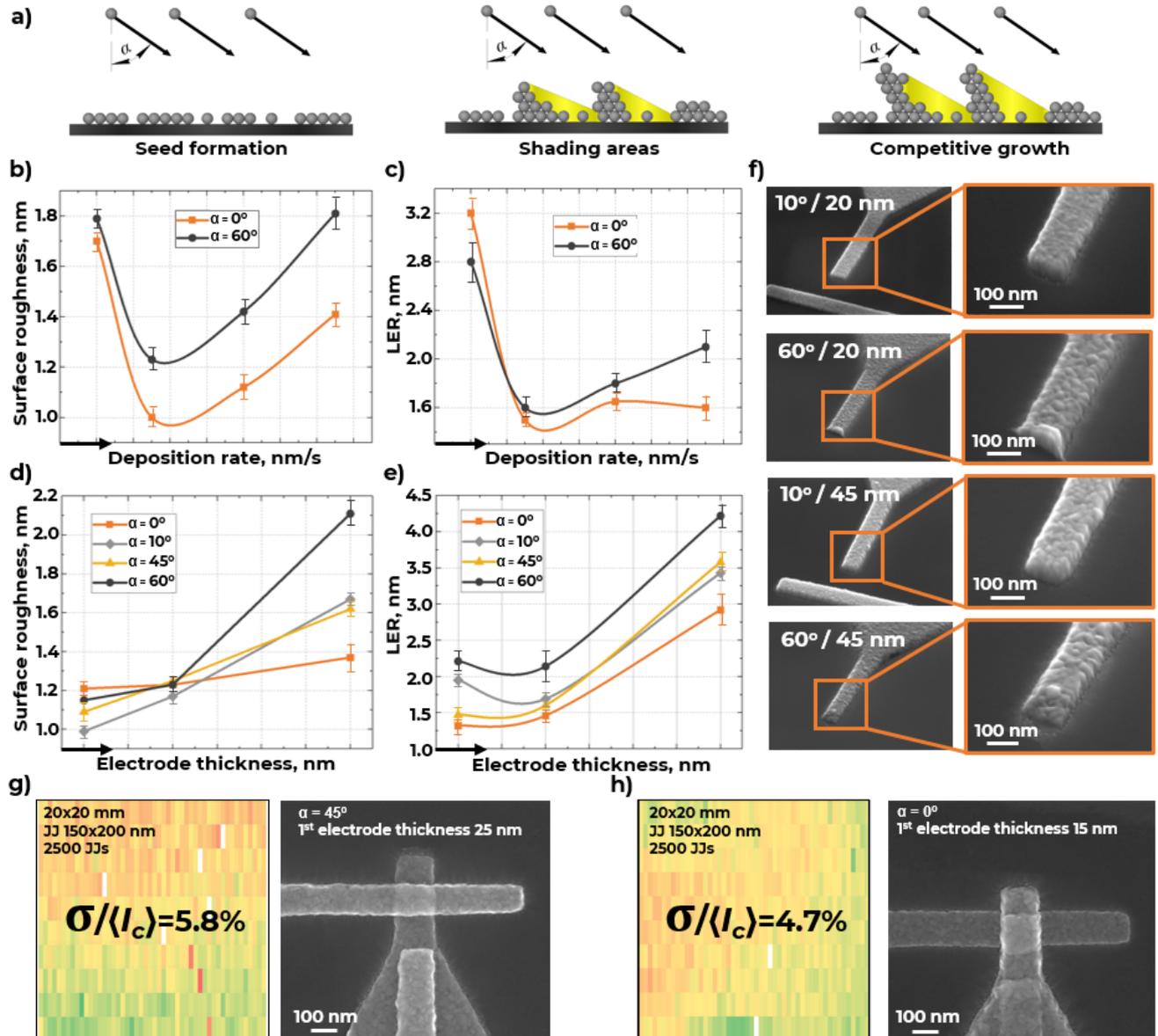

**Figure 2.** Angle shadow evaporation of Josephson junction bottom electrode: (a) schema of competitive islands growth during angled evaporation of electrodes; (b) surface roughness (RMS) and (c) line edge roughness of bottom JJ electrode vs deposition rate and evaporation angle; (d) surface roughness (RMS) and (e) line edge roughness of bottom JJ electrode vs its thickness and evaporation angle; (f) SEM image of the bottom JJ electrodes surface with various evaporation angles and thicknesses; critical current variation ($\sigma/\langle I_c \rangle$) of JJs evaporated at (g) 25 nm/45° and (h) 15 nm/0° deposition scheme.

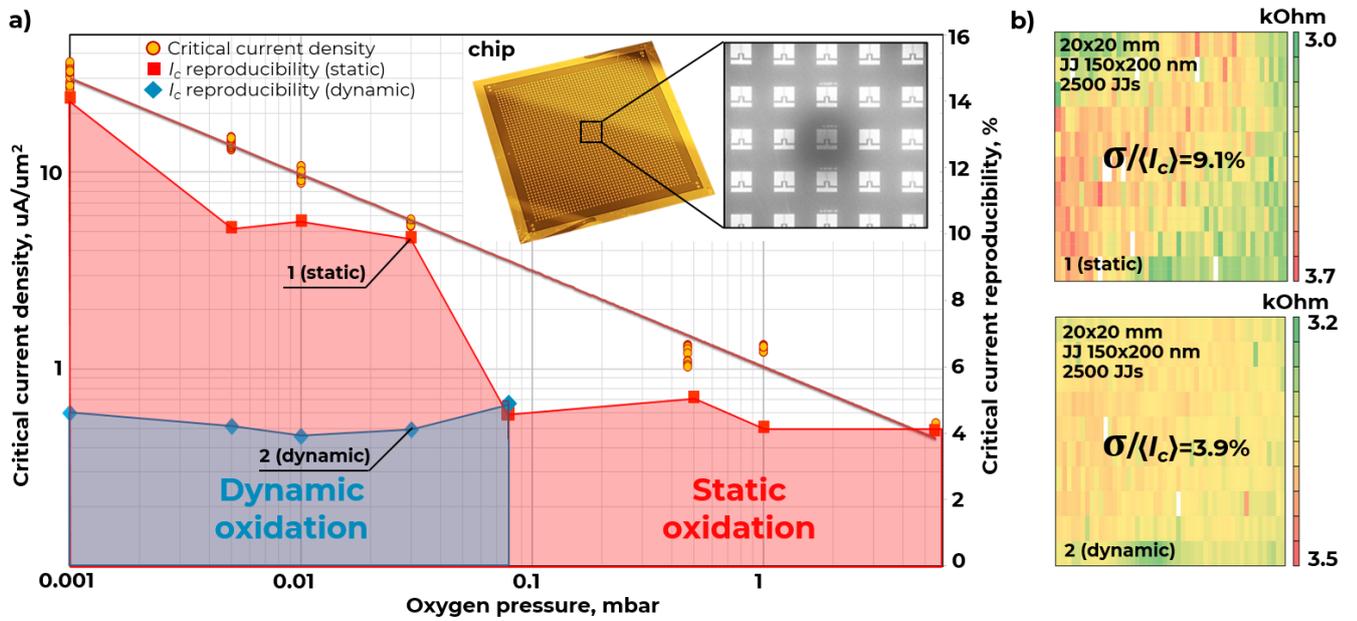

**Figure 3.** Critical current density and critical current reproducibility of the Al/AlO$_x$/Al junctions vs oxidation method and oxygen pressure (a); critical current variation ($\sigma/\langle I_c \rangle$) of JJs for static and dynamic oxidation (b).

One can use the combination of small deposition angle and thickness to fabricate high quality (RMS ~ 1 nm, LER < 1.3 nm) bottom JJ electrode (Fig. 2f).

We fabricated two test 20×20 mm$^2$ area chips to evaluate the influence of the surface roughness of thin film bottom electrodes on JJ critical current reproducibility (determined from the normal-state resistance [22]). Each chip contained more than 2500 JJs with different areas from 0.01 to 0.09 um$^2$. The first chip was fabricated with 25 nm thick JJs bottom electrodes evaporated at 45° angle; the top electrode was deposited orthogonally. The second chip contained JJs with 15 nm thick bottom electrodes deposited orthogonally (0°); the top electrode was deposited at 45°. One can notice, that 15 nm/0° deposition scheme ensured reducing the spread of the critical current ($\sigma/\langle I_c \rangle$) for 150x200 nm$^2$ area JJs (widely used in qubits design) over the chip from 5.8 % to 4.7 % (Fig. 2g,h). Critical current variation maps over the chips for the JJ with different areas can be found in the supplementary materials.

**Oxidation of Josephson junctions.** Josephson junctions critical current is determined by thickness and structure of the tunnel barrier [23]. It is formed at the stage of the bottom metal electrode oxidation. Oxidation process parameters like oxygen pressure and time (exposure time in O$_2$), oxidation method (static/dynamic) have dramatic influence on tunnel barrier properties [24]. Direct control of tunnel barriers is quite difficult, because its size is varying from 100 nm to 1 µm at 1-2 nm barrier thickness [25]. Usually, the tunnel barrier is characterized by electrical measurements.

First, we investigated the effect of oxygen pressure and oxidation time on the JJ critical current density ($j_c$) which is independent on the junction area. Fig. 3a demonstrates the experimental dependence of the critical current density vs oxygen pressure in the range from 0.001 to 10 mbar. One can see that the obtained $j_c$ range makes it possible to fabricate JJs in a wide range of critical currents required both for qubit circuits and for Josephson parametric amplifiers.

Next, we experimentally showed the influence of the oxidation method on the critical current variation. During the study, an increase in the critical current variation across the chip was observed using the static oxidation method at low oxidation pressures (less than 0.1 mbar). It can be explained by oxygen leaking into the chamber during the oxidation process at very low pressures [23]. Transition to dynamic oxidation allowed reducing the spread of critical current over a chip for the mentioned above range of oxygen pressures. According to resistance measurements at room temperature for JJs size of 150×200 nm, it is possible to reduce the spread ($\sigma/\langle I_c \rangle$) to 3.9 % across the chip using dynamic oxidation (Fig. 3b). Critical current variation maps over the chips for the JJ with different areas can be found in the supplementary materials.

**Qubit frequency repeatability.** We fabricated three similar chips with 6 transmon qubits on each to correlate observed improvements in the Josephson junction uniformity with device performance. The six fixed-frequency transmons were designed for target frequency of 4.3 GHz with detunings between neighbors. Each chip was fabricated individually with the same optimized shadow evaporation and oxidation parameters for the Josephson junctions described in this work. Each chip was packaged in a magnetically- and infrared-shielded copper sample holder mounted at the base plate of a dilution refrigerator at temperature below 15 mK. Microwave attenuators were used to isolate the qubit chip from thermal and instrumental noise. The readout line is equipped with an impedance matched parametric amplifier (IMPA) [26]. At the stage (4K) of the cryostat, a high electron mobility transistor (HEMT) is installed. The output line is further amplified outside the cryostat with amplifier.

| Qubit | | $f_{01}$, GHz | $T_1$, us | $T_2^*$, us |
|---|---|---|---|---|
| *Chip #1* | #1 | 4.43 | 143.6 | 19.9 |
| | #2 | 4.41 | 148.0 | 49.3 |
| | #3 | 4.37 | 116.1 | 24.9 |
| | #4 | 4.29 | 132.3 | 43.9 |
| | #5 | 4.33 | 103.3 | 74.6 |
| | #6 | 4.45 | 115.4 | 28.8 |
| Average | | 4.38 | 126.45 | 40.23 |
| **st. deviation** | | **1.28 %** | **12.74 %** | **46.01 %** |
| *Chip #2* | #1 | 4.30 | 120.9 | 19.4 |
| | #2 | 4.24 | 138.8 | 21.9 |
| | #3 | 4.29 | 172.9 | 33.9 |
| | #4 | 4.16 | 117.5 | 10.8 |
| | #5 | 4.13 | 143.6 | 13.0 |
| | #6 | 4.26 | 109.9 | 43.0 |
| Average | | 4.23 | 133.93 | 23.67 |
| **st. deviation** | | **1.51 %** | **15.70 %** | **48.15 %** |
| *Chip #3* | #1 | 4.29 | 275.7 | 20.4 |
| | #2 | 4.28 | 250.9 | 5.3 |
| | #3 | 4.35 | 210.3 | 13.2 |
| | #4 | 4.23 | 152.2 | 5.8 |
| | #5 | 4.33 | 134.9 | 22.5 |
| | #6 | 4.35 | 138.2 | 26.8 |
| Average | | 4.31 | 193.70 | 14.60 |
| **st. deviation** | | **1.00 %** | **28.69 %** | **48.07 %** |
| Total aver. | | 4.31 | 151.36 | 26.17 |
| **Total st. dev.** | | **1.91 %** | **30.77 %** | **64.69 %** |

**Table 1.** Parameters of the measured qubits.

After identifying the qubit frequency using spectroscopy, we first perform a measurement of Rabi oscillations where an excitation pulse is applied at the qubit frequency, followed by a readout pulse on the resonator. From the Rabi oscillations, we extract the pulse length required to excite the qubit. Next, we use our calibrated π-pulse duration to perform a Ramsey experiment to determine coherence time. Finally, we extract the qubit relaxation time by applying a π-pulse and increase the delay between the qubit excitation and readout pulses. To reliably characterize the system, the pulse schemes are repeated and the measurement outputs are averaged.

As a result, we measured frequencies, relaxation ($T_1$) and coherence ($T_2^*$) time of each qubit on three different chips (Table 1). We observed high repeatability of qubits frequencies with a spread less than 1.5 % over the 5×10 mm$^2$ chip area and less than 2 % over three chips. The qubit measurement results for the chips described in this work are listed in the supplementary materials.

## Conclusions

Motivated by required ultrahigh uniformity of Josephson junctions critical current for scaling superconducting quantum circuits with multiple qubits, we undertook a study of optimal Josephson junction electrode thickness, deposition rate and deposition angle. We experimentally confirmed that decreased deposition angle and thin electrode contribute to RMS surface roughness and electrode line edge roughness minimization (RMS ~ 1 nm, LER < 1.3 nm). In addition, the influence of oxidation method, pressure and time on critical current reproducibility was determined. We demonstrated that transition to dynamic oxidation allowed reducing the spread of critical current over a chip for oxygen pressures less than 0.1 mbar.

Utilizing the proposed method, we achieved Al/AlO$_x$/Al junction critical current variation ($\sigma/\langle I_c \rangle$) less than 3.9 % (for the JJs from 150×200 to 150×600 nm$^2$) and 7.7 % (for 100×100 nm$^2$) over 20×20 mm$^2$ chip. Finally, we fabricate separately three 5×10 mm$^2$ chips with 18 transmon qubits showing less than 1.9 % frequency variation between qubits on different chips. The proposed approach and optimization criteria can be utilized for a robust wafer-scale superconducting qubit circuits fabrication.

## References

1. Clarke, J., & Wilhelm, F. K. Superconducting quantum bits. Nature **453**, 7198, 1031-1042. https://doi.org/10.1038/nature07128 (2008).
2. Zeng, L., Tran, D., Tai, CW. et al. Atomic structure and oxygen deficiency of the ultrathin aluminium oxide barrier in Al/AlOx/Al Josephson junctions. Sci Rep **6**, 29679. https://doi.org/10.1038/srep29679 (2016).
3. Krantz, P., Kjaergaard, M., Yan, F., Orlando, T. P., Gustavsson, S., & Oliver, W. D. A quantum engineer's guide to superconducting qubits. Appl. Phys. Rev. **6**, 021318. https://doi.org/10.1063/1.5089550 (2019).

4. Besedin, I. S. et al. Topological excitations and bound photon pairs in a superconducting quantum metamaterial. Phys. Rev. B **103**, 224520. https://doi.org/10.1103/PhysRevB.103.224520 (2021).
5. Fedorov, G. P. et al. Photon transport in a Bose-Hubbard chain of superconducting artificial atoms. Phys. Rev. Lett. **126**, 180503. https://doi.org/10.1103/PhysRevLett.126.180503 (2021).
6. Moskalenko, I.N., Simakov, I.A., Abramov, N.N. et al. High fidelity two-qubit gates on fluxoniums using a tunable coupler. npj Quantum Inf **8**, 130. https://doi.org/10.1038/s41534-022-00644-x (2022).
7. Kreikebaum, J. M., O'Brien, K. P., Morvan, A., & Siddiqi, I. Improving wafer-scale Josephson junction resistance variation in superconducting quantum coherent circuits. Supercond. Sci. Technol. **33** 06LT02. https://doi.org/10.1088/1361-6668/ab8617 (2020).
8. Osman, A., et al. Simplified Josephson-junction fabrication process for reproducibly high-performance superconducting qubits. Appl. Phys. Lett. **118**, 064002. https://doi.org/10.1063/5.0037093 (2021).
9. Verjauw, J., Acharya, R., Van Damme, J. et al. Path toward manufacturable superconducting qubits with relaxation times exceeding 0.1 ms. npj Quantum Inf **8**, 93. https://doi.org/10.1038/s41534-022-00600-9 (2022).
10. Pop, I. M. et al. Fabrication of stable and reproducible submicron tunnel junctions. Journal of Vacuum Science & Technology B **30**, 010607. https://doi.org/10.1116/1.3673790 (2012).
11. Macklin, C. et al. A near-quantum-limited Josephson traveling-wave parametric amplifier. Science **350**.6258, 307-310. https://doi.org/10.1126/science.aaa8525 (2015).
12. Zorin, A. B. Josephson traveling-wave parametric amplifier with three-wave mixing. Phys. Rev. Applied **6**, 034006. https://doi.org/10.1103/PhysRevApplied.6.034006 (2016).
13. Rodionov, I.A., Baburin, A.S., Gabidullin, A.R. et al. Quantum Engineering of Atomically Smooth Single-Crystalline Silver Films. Sci Rep **9**, 12232. https://doi.org/10.1038/s41598-019-48508-312232 (2019).
14. Dolan G. J. Offset masks for lift-off photoprocessing. Appl. Phys. Lett. **31**.5, 337. https://doi.org/10.1063/1.89690 (1977).
15. Douglass Jr, D. H., & Meservey, R. Energy gap measurements by tunneling between superconducting films. I. Temperature dependence. Phys. Rev. **135**, A19. https://doi.org/10.1103/PhysRev.135.A19 (1964).
16. Zeng, L. J. et al. Direct observation of the thickness distribution of ultra thin AlOx barriers in Al/AlOx/Al Josephson junctions. J. Phys. D: Appl. Phys. **48** 395308. https://doi.org/10.1088/0022-3727/48/39/395308 (2015).
17. Nik, S. et al. Correlation between Al grain size, grain boundary grooves and local variations in oxide barrier thickness of Al/AlOx/Al tunnel junctions by transmission electron microscopy. SpringerPlus **5**.1. https://doi.org/10.1186/s40064-016-2418-8 (2016).
18. Fritz, S., Seiler, A., Radtke, L. et al. Correlating the nanostructure of Al-oxide with deposition conditions and dielectric contributions of two-level systems in perspective of superconducting quantum circuits. Sci Rep **8**, 7956. https://doi.org/10.1038/s41598-018-26066-4 (2018).
19. Barranco, A. et al. Perspectives on oblique angle deposition of thin films: From fundamentals to devices. Progress in Materials Science **76**, 59-153. https://doi.org/10.1016/j.pmatsci.2015.06.003 (2016).
20. Shen, D. et al. Character and fabrication of Al/Al2O3/Al tunnel junctions for qubit application. Chinese Science Bulletin **57**.4, 409-412. https://doi.org/10.1007/s11434-011-4821-4 (2012).
21. Dirks, A. G. and Leamy, H. J. Columnar microstructure in vapor-deposited thin films. Thin solid films **47**.3, 219-233. https://doi.org/10.1016/0040-6090(77)90037-2 (1977).
22. Ambegaokar, V. and Baratoff, A. Tunneling between superconductors. Phys. Rev. Lett. **10**, 486. https://doi.org/10.1103/PhysRevLett.10.486 (1963).
23. Wu, Y. L. et al. Fabrication of Al/AlOx/Al Josephson junctions and superconducting quantum circuits by shadow evaporation and a dynamic oxidation process. Chinese Phys. B **22** 060309. https://doi.org/10.1088/1674-1056/22/6/060309 (2013).
24. Holmes, D. S., & McHenry, J. (2016). Non-normal critical current distributions in Josephson junctions with aluminum oxide barriers. IEEE Transactions on Applied Superconductivity, **27**.4, 1-5. https://doi.org/10.1109/TASC.2016.2642053 (2017).
25. Zeng, L., Tran, D., Tai, CW. et al. Atomic structure and oxygen deficiency of the ultrathin aluminium oxide barrier in Al/AlOx/Al Josephson junctions. Sci Rep **6**, 29679. https://doi.org/10.1038/srep29679 (2016).
26. Ezenkova, D., et al. Broadband SNAIL parametric amplifier with microstrip impedance transformer. Appl. Phys. Lett. **121**, 232601. https://doi.org/10.1063/5.0129862 (2022).

**Acknowledgements**

**Author contributions statement**

**Additional information**